\documentclass[%
 reprint,
 amsmath,amssymb,
 aps,
]{revtex4-2}

\usepackage{graphicx}
\usepackage{dcolumn}
\usepackage{bm}
\usepackage{xcolor}
\usepackage{enumitem}
\usepackage{soul}

\begin{document}

\preprint{APS/123-QED}

\title{Generating persistent-current superpositions in Bose-Einstein condensates using dynamic optical potentials}

\author{Renzo Testa}
\affiliation{SUPA School of Physics $\&$ Astronomy, University of St Andrews, North Haugh, St Andrews KY16 9SS, UK}

\author{Donatella Cassettari}
\affiliation{SUPA School of Physics $\&$ Astronomy, University of St Andrews, North Haugh, St Andrews KY16 9SS, UK}

\date{\today}

\begin{abstract}
    Precise and flexible manipulation of the motional state of ultracold atoms is a fundamental enabling technology for diverse applications such as quantum sensing and quantum computation. In this paper we propose a general, simple and highly efficient method to engineer the motional state of a Bose-Einstein condensate with time-dependent optical fields, which can be realized experimentally with existing light sculpting techniques. We demonstrate numerically how to engineer superpositions of persistent currents in a toroidal trap, achieving very high fidelity. We also study in detail the stability of the state over time, and we present an analytical two-state model that approximates well the evolution of the state in presence of self-interactions.  
\end{abstract}

\maketitle


\section{\label{sec:intro}introduction}

In recent years, remarkable progress has been made in controlling the motional state of ultracold atoms in traps and circuits. This led to the emerging field of atomtronics \cite{Amico}, seeking to realize circuits in which ultracold atoms are manipulated in versatile optical and magnetic guides. In particular, persistent currents in toroidal traps are of fundamental interest in quantum research, as well as forming the basis for atomtronic quantum devices \cite{Polo_review_article}. Methods to induce persistent currents include stirring the gas with a laser beam \cite{Wright_phase_slips,Cai_PersistentCurrents}, two-photon Raman transitions \cite{Ramanathan_SuperflowToroidalBEC, Moulder_supercurrent_decay, Phillips_persistent_flow}, and phase imprint \cite{Roati_persistent_currents,Perrin}. Persistent currents have been used in many scenarios, for instance in the study of turbulence \cite{Neely_turbolence, Roati_vortex_instabilities}, in the quenching of Fermi gases \cite{Allman_quenching}, and to verify the geodesic rule in stochastic currents by merging independent condensates \cite{Beugnon_bec_merging}. There are many proposals building on these experimental results, for instance on how the transfer of angular momentum between adjacent rings can be used to sense accelerations \cite{Chaika_atomtronics}.

\begin{figure*}[t]
\includegraphics[scale=0.52]{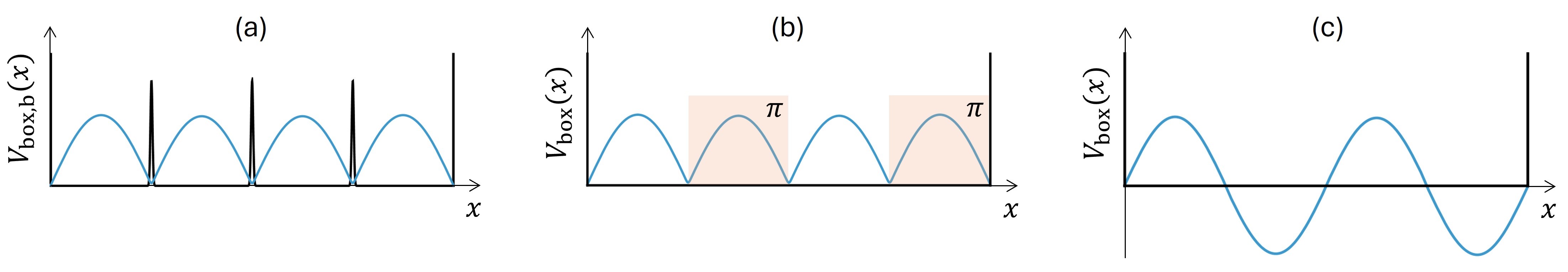}
\vspace{-0.2cm}
\caption{\label{fig:Scheme_BarrierPI} Wave function engineering in a linear trap $V_\text{box}(x)$. (a) The initial state is the ground state of potential $V_\text{box,b}(x)$. (b) We suddenly remove the barriers and, at the same time, apply a $\pi$-phase imprint to the $2^\text{nd}$ and $4^\text{th}$ lobes of the wave function. (c) Final state: the phase imprint flips the corresponding parts of the wave function, so that the resulting state closely approximates the $3^\text{rd}$ excited state of $V_\text{box}(x)$.}
\end{figure*}

Another important frontier is the realization of superpositions of persistent currents. These will find applications in quantum information processing \cite{Dowling_VortexPhaseQubit, Dowling_OAM}, and in quantum sensing \cite{Dowling_MatterWaveGyroscopy, Pelegri_UnbalancedVortices}. A superposition of persistent currents can be seen as a type of guided atom interferometer in which the two interfering waves are uniformly spread around the ring, instead of originating from a localized wave packet which is then split and recombined. Assuming the current superposition is long lived, this interferometer can be used to measure rotations \cite{Dowling_MatterWaveGyroscopy} or magnetic fields \cite{Pelegri_UnbalancedVortices} with high sensitivity. In general, guided atom interferometers \cite{Prentiss_rot_sensing, Stamper_bec_interferometry, Garrido_gyrometers, Sackett_Sagnac, Baker_rotation_sensing, Boshier_gyroscope} have the advantage of being more compact and portable compared to free-space interferometers. However, superpositions of persistent currents have not yet been experimentally demonstrated. Superpositions of vortices in simply-connected traps have been demonstrated \cite{Bigelow_vortex_sculting, Wuhan2024_Vortices}, but vortices are in general much shorter lived compared to persistent currents in ring traps.

In this paper we propose a general method to transfer a Bose-Einstein condensate to motional states that are in principle arbitrary, and we show in particular its application to the case of superpositions of persistent currents. The main idea is that the condensate wave function has an amplitude and a phase, each of which can be controlled independently to create a given target state: the amplitude can be controlled by shaping the trapping potential, and the phase by applying a phase imprint. This gives full freedom to engineer the wave function. To this purpose, it is possible to use arbitrary, reconfigurable optical fields enabled by devices such as fast-scanning acousto-optic deflectors, digital micromirror devices, and liquid-crystal spatial light modulators \cite{Gauthier_HiResTrapping}. Specifically, time-dependent optical fields can be used both to shape the trapping potential and to imprint a phase. The latter is achieved with a pulsed optical field, as was first used to generate dark solitons \cite{Burger_dark_solitions, Phillips_solitons}, and vortices \cite{Ertmer_vortices}. Hence, the entire sequence of shaping the trapping potential and imprinting a phase only necessitates spatio-temporal control of the light intensity.

The paper is organized as follows. In Sec.~\ref{sec:wave_function_engineering} we introduce the principle of our wave function engineering protocol and its application to the case of superpositions of persistent currents. Sec. \ref{sec:numerical_model} describes the methodology used in our computational work, while Secs. \ref{sec:Initial_fidelity} and \ref{sec:Stability} show the fidelity and stability of the engineered states. Finally, Sec. \ref{sec:conclusions} provides a summary and a perspective of possible future developments. The appendices report further numerical analysis supporting our results, as well as an analytical two-state model that captures how superpositions of persistent currents evolve in time.

\section{\label{sec:wave_function_engineering}wave function engineering}

To introduce the principle of wave function engineering, we first consider a condensate in a linear trap in the limit of no self-interactions. So far, control of the vibrational state in a linear trap has been investigated with two methods: gradual trap deformation, and the application of a time-dependent perturbation (e.g. “shaking”). For instance in \cite{Garaot_VibrationalMultiplexing,Garaot_FastDriving}, the transfer from the ground state to an excited state of the trap was simulated by means of gradual trap deformations. In \cite{Bucker_StateInversion,Frank_InterferometryMotStates}, the transfer to the first excited state of a trap was implemented experimentally by “shaking” the trap along a trajectory calculated with optimal control theory. 

The approach we propose here is an alternative to these methods. We control both the phase and the amplitude of the condensate wave function with a combination of time-varying trapping potentials and phase imprint. This approach offers freedom to engineer the wave function and to transfer atoms to arbitrary excited states in the trap. To illustrate this point, Fig. \ref{fig:Scheme_BarrierPI} describes how we populate a given target state, here the $3^\text{rd}$ excited state, in a box potential $V_\text{box}(x)$:

\begin{enumerate}[itemsep=1pt,parsep=0pt,topsep=3pt]
\item We start with atoms in the ground state of the $V_\text{box,b}(x)$ potential, which is $V_\text{box}(x)$ with added barriers at the locations where the target state has nodes.
\item Next, we suddenly remove the barriers. At the same time, we apply a phase imprint of $\pi$ to the $2^\text{nd}$ and $4^\text{th}$ lobes of the wave function. This phase imprint flips these parts of the wave function, so that the resulting state closely reproduces the target state.
\end{enumerate}

In practice, the barriers have finite height and width, which is diffraction limited. This leads to a discrepancy between the final state and the target state. However, the numerical implementation in Fig.~\ref{fig:Scheme_BarrierPI} shows this discrepancy to be small. A finite height of the barrier is actually a necessary feature, as sufficient tunneling is required to avoid fragmentation and to ensure a uniform phase across the different lobes before applying the phase imprint. 

\begin{figure}[h!]
\includegraphics[scale=0.22]{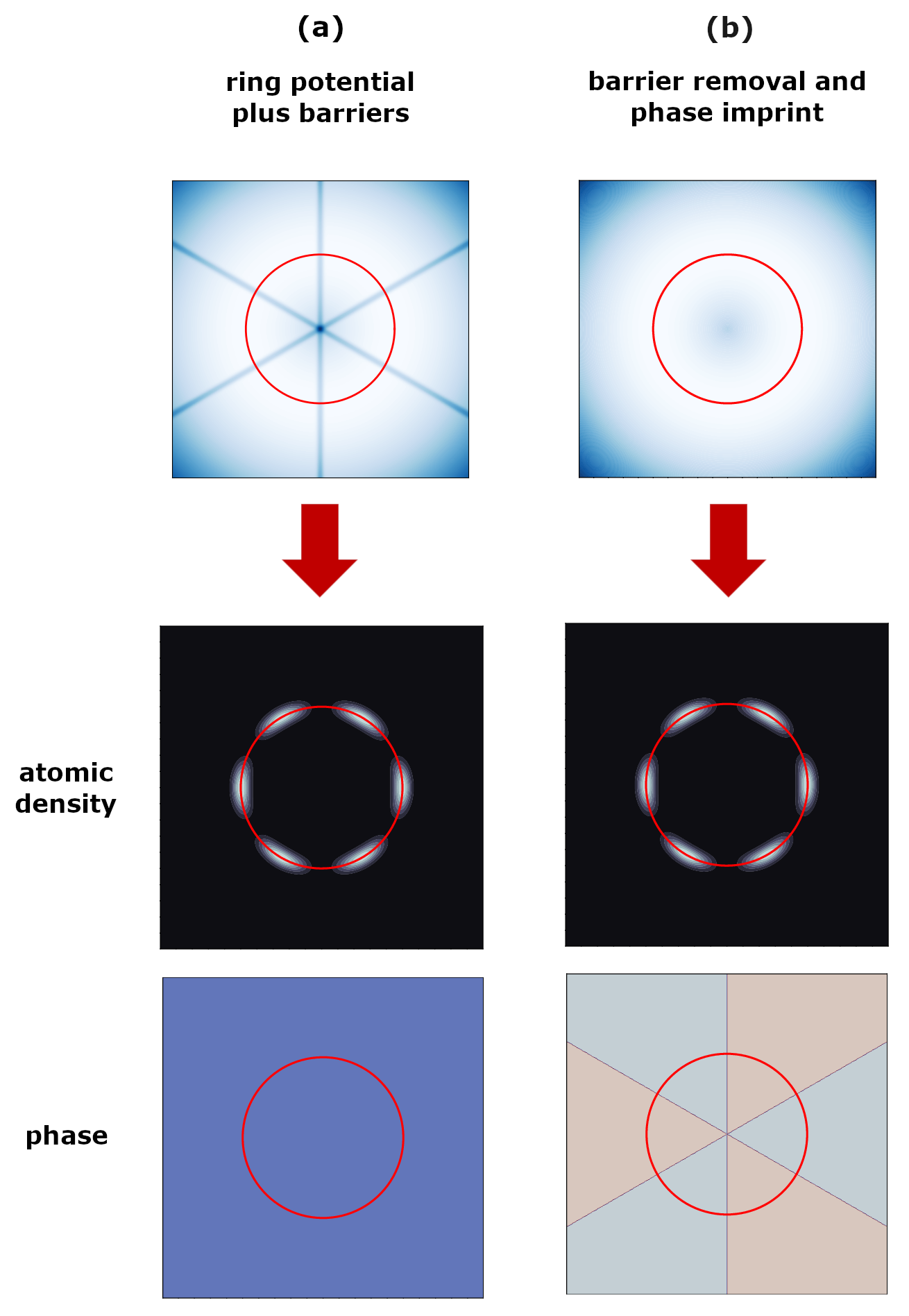}
\caption{\label{fig:Ring_BarrierPI} Wave function engineering in a ring trap. The red circle indicates the ring radius. The barriers are created with linear repulsive potentials. In this case, we populate an $|\textrm{OAM} \rangle$ state with $m=3$.}
\end{figure}

Next, we focus on the motional states of atoms in 2D circuits. In ring geometries, this approach can be used to generate superpositions of persistent currents. Starting from a single persistent current state $|m \rangle=\frac{1}{\sqrt{2\pi}}f(r)e^{\textrm{i}m\phi}$, where $m$ is the angular momentum quantum number, $\phi$ is the azimuthal angle around the ring and $f(r)$ is the radial part of the wave function (with $\int f^2(r)rdr = 1$), we define our target state  $|\textrm{OAM} \rangle$ as the superposition: 
\begin{equation}
    \label{eq:OAM_cos}
    |\textrm{OAM} \rangle=\frac{1}{\sqrt{2}}(|m \rangle + |-m \rangle) =\frac{1}{\sqrt{\pi}}f(r)\cos(m\phi).
\end{equation} 
The resulting atomic density goes as $\cos^2(m \phi)$, resulting in $2m$ nodes. 

The protocol for populating the $|\textrm{OAM} \rangle$ target state is illustrated in Fig. \ref{fig:Ring_BarrierPI} for $m=3$. Similarly to the case of the excited state in the linear trap, here we consider the ring trap to which $m$ linear repulsive potentials have been added. Each repulsive potential intercepts the ring on two diametrically opposite points, creating $2m$ barriers around the ring at angles equispaced by $180^{\circ}/ m$. Starting with atoms in the ground state of this potential, we suddenly remove the barriers while at the same time applying a phase imprint of $\pi$ to alternate sectors. The resulting state, which we indicate as $|\textrm{ENG} \rangle$, reproduces closely the $\cos(m\phi)$ structure of $|\textrm{OAM} \rangle$ in Eq. \ref{eq:OAM_cos}. 

The comparison between $|\textrm{ENG} \rangle$ and $|\textrm{OAM} \rangle$ at the time of barrier removal and phase imprint can be formulated in terms of the fidelity $F$:
\begin{equation}
    \label{eq_initial_fidelity}
    F = |\langle \textrm{OAM}| \textrm{ENG} \rangle|^2
\end{equation}
Here $F=1$ means a perfect reconstruction of the target state. As in the previous example of the linear trap, the fidelity is determined by the parameters of the barriers. We anticipate that, even though we have considered a non-interacting condensate so far, in the following we find good fidelities also in the case of a weakly interacting condensate. 

Our protocol is an alternative to the method proposed in \cite{Dowling_VortexPhaseQubit} and implemented in \cite{Wuhan2024_Vortices}, in which the $|\textrm{OAM} \rangle$ state is created by a superposition of Laguerre-Gaussian beams transferring angular momentum to a condensate via a two-photon Raman transition. The LG superposition is created by superimposing two beams of opposite vorticity, and this overlap leads to an interference pattern in the optical field characterized by bright lobes along the azimuthal direction. As a consequence of this, the two-photon Rabi frequency is space dependent, which gives a theoretical limit to the transfer efficiency as shown in \cite{Wuhan2024_Vortices}. Specifically, there is an optimal two-photon pulse duration that gives a transfer efficiency of slightly less than $50\%$. The remaining population is left in the initial internal state, which is non-rotating. In comparison, our method of barrier and phase imprint ensures high efficiency, while maintaining high fidelity $F$ as shown in Section \ref{sec:Initial_fidelity}.

\section{\label{sec:numerical_model}Numerical model}
Our simulations require the numerical integration of the 2D Gross-Pitaevski equation on the $x-y$ plane:
\begin{equation}
    \label{Eq_GrossPitaevsi}
    \textrm{i}\hbar\frac{\partial\psi}{\partial t}=\left[  - \frac{\hbar^2}{2M} \nabla^2 +V(x,y) + g_{2D}|\psi|^2 \right] \psi 
\end{equation}
The mean-field constant is $g_{2D}=(N\sqrt{8\pi} \hbar^2 a_s) / (M a_z)$, where $N$ is the number of atoms in the condensate, $a_s$ and $M$ are the scattering length and the atomic mass respectively, and $a_z = \sqrt{\hbar/(M \omega_z)}$ is the harmonic oscillator length along the transversal $z$ direction \cite{Pitaevski_Stringari_Book}. The physical constants are those of $^{87}\textrm{Rb}$, which offers relatively weak repulsive self-interactions. We focus on regimes of fairly small atom numbers, namely $N=10^3$ and $N=10^4$. As a benchmark, to help with the interpretation of the results in presence of self-interactions, we also consider the non-interacting case by putting $g_{2D}=0$.

Our ring trap has a radius of $r_0$ = 50 $\mu\text{m}$, with the radial confinement given by the potential:
\begin{equation}
    V_{\text{ring}}(r) = \frac{1}{2}M\omega_{\text{trap}}^2 (r - r_0)^2.
\end{equation}
The trapping frequency is set via the harmonic oscillator length in the $x-y$ plane $a_{\text{ho}} =\sqrt{\hbar/(M\omega_{\textrm{trap}})}$, by setting $a_{\text{ho}}=r_0/10$, which gives $\omega_{\text{trap}} \approx 29$ rad/s. The trapping frequency $\omega_z$ along the transversal $z$-axis is $4$ times higher than $\omega_{\text{trap}}$. This overall weak confinement, in conjuction with relatively small atom numbers, leads to a small mean-field energy which favors the stability of the current superposition. The mean-field energy is always less than $\hbar\times$30~rad/s, which is well below $\hbar\omega_z$, hence satisfying the 2D criterion \cite{Pitaevski_Stringari_Book}.

The integration of Eq. \ref{Eq_GrossPitaevsi} is performed using the Trotter-Suzuki software package \cite{TrotterSuzuki}. The spatial discretization is 401 grid points over a region of $4r_0$, which allows to place a few points within the barrier extension. We have verified that our results are unchanged if we double the spatial discretization.

First, we calculate the $|\textrm{OAM} \rangle$ state in the ring trap. For this, we use imaginary time evolution to find the ground state of the ring trap, to which we then apply the superposition of phase windings $(e^{\textrm{i}m\phi} + e^{-\textrm{i}m\phi})$. 

To create the $|\textrm{ENG} \rangle$ state, we find the ground state of the ring trap with added barriers with imaginary time evolution. Then, to implement the protocol outlined in the previous section, we suddenly remove the barriers and apply the phase imprint of $\pi$ to alternate sectors. Finally, the $|\textrm{ENG} \rangle$ state so constructed is compared with $|\textrm{OAM} \rangle$ by calculating the fidelity $F$ as in Eq. \ref{eq_initial_fidelity}. 

The barriers are chosen to be thin Gaussians. Their width is kept fixed at a value close to the diffraction limit of a typical optical system, as larger widths would reduce the fidelity. In particular, we choose Gaussians with FWHM=3.3~$\mu$m, which is close to what has been experimentally realized with a digital micromirror device \cite{Roati_supercurrents}.

We consider two cases, $m=3$ and $m=9$. The barrier locations are equispaced, respectively, by $60^{\circ}$ and $20^{\circ}$. Having fixed the barrier width, first we optimize the fidelity as a function of the barrier height $h_{\textrm{barrier}}$, as shown in Sec. \ref{sec:Initial_fidelity}. Subsequently, we study the time evolution to demonstrate the stability of the states through their autocorrelation function. For this we integrate Eq.~\ref{Eq_GrossPitaevsi} in real time with a time step of $5\times10^{-6}$~s. We record the evolution of the system dynamics at intervals of 500 steps corresponding to time frames of $2.5$~ms. The results are shown in Sec. \ref{sec:Stability}.

\begin{figure}[h!]
\includegraphics[scale=0.48]{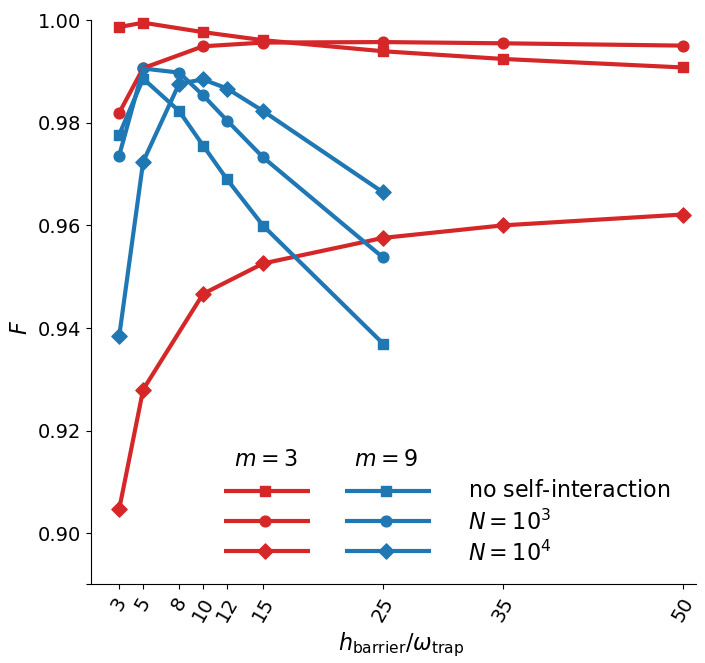}
\caption{\label{fig:Fidelity0} Fidelity $F = |\langle \textrm{OAM}| \textrm{ENG} \rangle|^2$ as a function of barrier height for a non-interacting condensate, and for interacting condensates with $N=10^3,10^4$. The two cases $m=3$ and $m=9$ are shown.}
\end{figure}

\section{\label{sec:Initial_fidelity}Fidelity optimization}

Given that our goal is for the $|\textrm{ENG} \rangle$ state to be as close as possible to the target $|\textrm{OAM} \rangle$ state, we characterize the fidelity $F$ between the two states as a function of $h_{\textrm{barrier}}$. As shown in Fig. \ref{fig:Fidelity0}, we stay below $h_{\textrm{barrier}}/\omega_{\textrm{trap}}=50$, which corresponds to $k_B\times$11 nK. This range of heights is limited by the requirement of uniform phase across the different lobes of the wave function, which is achieved in the tunneling regime. Barriers that are too high would lead to fragmentation, i.e. uncorrelated condensates in the different segments of trap with uncontrolled phase differences between them, undermining the protocol implementation. Studies of the transition from phase coherence to fragmentation \cite{Japha_coherence,Spekkens_fragmentation} confirm that we can assume coherence within our range of barrier heights.

Fig. \ref{fig:Fidelity0} shows that we reach high $F$ values ($>90\%$) in all the cases we investigated. In most cases, we find an optimal value of $h_{\textrm{barrier}}$: starting from very low $h_{\textrm{barrier}}$, the fidelity always increases because the presence of the barrier locally reduces the value of the wave function, bringing it closer to the node of the $|\textrm{OAM}\rangle$ state. But if $h_{\textrm{barrier}}$ is too high, the wave function is displaced away from the barrier, leading to lobes that are too separated compared to the $|\textrm{OAM}\rangle$ state. This effect is illustrated in Fig. \ref{fig:Displacement_NOINT_m3} of Appendix A. The displacement is localized to the region of the barrier width but its effect can distort the wave function much further away from the barriers. 

Fig. \ref{fig:Fidelity0} also displays a decay of the fidelity at larger $h_{\textrm{barrier}}$ for $m=9$, leading to a better defined optimal value of $h_{\textrm{barrier}}$ compared to the $m=3$ case. This is due to the smaller lobe size for higher $m$, meaning that the same displacement away from the barrier has a larger impact on the fidelity.

In the presence of self-interactions, the number of atoms also plays a role, with larger $N$ offering lower fidelities. This is because the repulsive interactions widen the lobes of the $|\textrm{ENG}\rangle$ state, increasing the discrepancy from the $\cos^2(m \phi)$ structure of the $|\textrm{OAM}\rangle$ state. Another effect of self-interactions is that the atoms tend to fill the region of the barrier and so the optimal $h_{\textrm{barrier}}$ increases at larger $N$. This more clearly visible for $m=9$, which has well defined optimal $h_{\textrm{barrier}}$ values. In this case, the curve in
Fig. \ref{fig:Fidelity0} shifts to the right. In comparison, for $m=3$ and $N=10^4$ the optimal value is not reached in the investigated range.

\section{\label{sec:Stability}Stability}
We monitor the time evolution of the wave function for up to a few seconds after the phase imprint. We analyze how similar the state remains to itself by computing the square modulus of the autocorrelation function of $|\psi(t) \rangle$ at time $t$  with $|\psi(0) \rangle$ at $t=0$:
\begin{equation}
    R(t) = |\langle \psi(t) | \psi(0) \rangle|^2
\end{equation}
where $| \psi \rangle$ is either the $| \textrm{ENG} \rangle$ or $| \textrm{OAM} \rangle$ state. For each case we investigate, we use the optimal value of $h_{\textrm{barrier}}$ that maximizes $F$ in Fig. \ref{fig:Fidelity0}.

\begin{figure}[h!]
\includegraphics[scale=0.7]{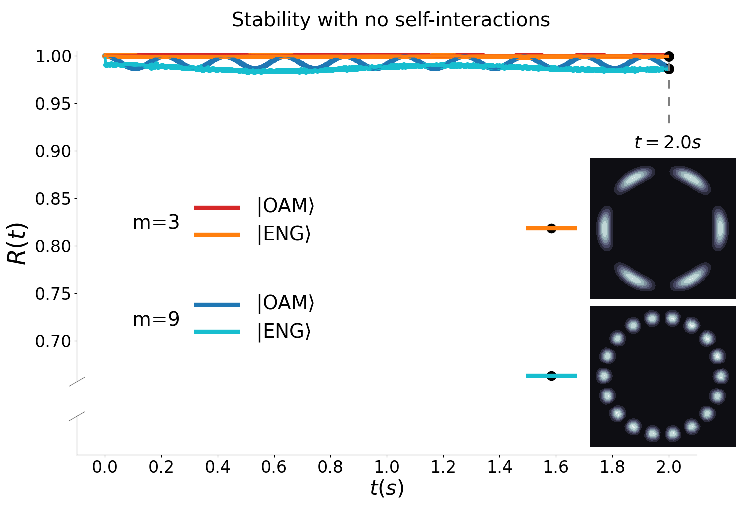}
\caption{\label{fig:Stability_NOINT} Square modulus of the autocorrelation function for the $|\textrm{OAM} \rangle$ and $|\textrm{ENG} \rangle$ states with no self-interactions. The two cases $m=3$ and $m=9$ are shown. The insets represent the atomic density for the $| \textrm{ENG} \rangle$ states with $m=3,9$ at $t=2$~s.}
\end{figure}

The non-interacting $|\textrm{OAM} \rangle$ state is a stationary state, i.e. it is an eigenstate of the Hamiltonian. This is confirmed by the numerical results shown in Fig. \ref{fig:Stability_NOINT}, where we see an almost stationary evolution, with practically constant $R(t)$. Only for the $|\textrm{OAM} \rangle$ state with $m=9$, we observe small oscillations. This is because imparting angular momentum to the condensate adds a centrifugal term to $V_{\text{ring}}(r)$, which shifts the minimum to a value larger than $r_0$. Hence the condensate finds itself out of equilibrium radially, and performs small oscillations around the new minimum with period $2 \pi/\omega_{\text{trap}}\approx 0.22$~s. This period is in perfect agreement with the oscillation of $R(t)$ in Fig.~\ref{fig:Stability_NOINT}.

\begin{figure*}[htbp!]
\includegraphics[scale=1.5]{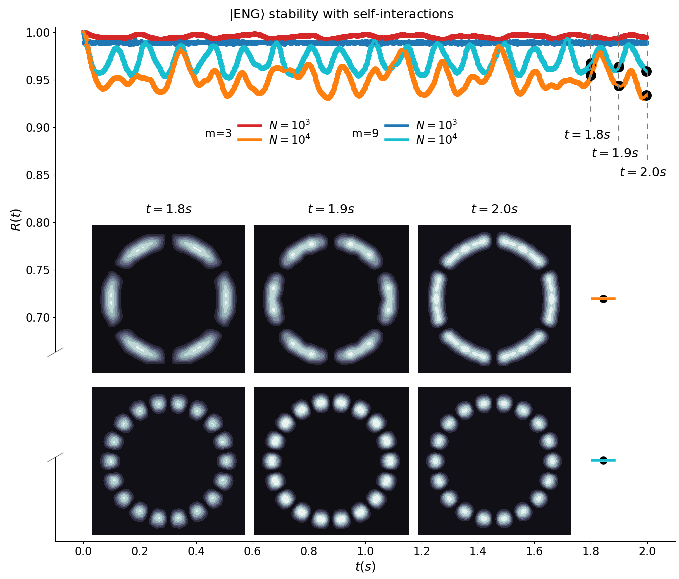}
\caption{\label{fig:Stability_INT}Square modulus of the autocorrelation function for the $|\textrm{ENG} \rangle$ state in presence of self-interactions, for $m = 3,9$ and $N = 10^3, 10^4$. The insets represent the atomic density at chosen times for $N=10^4$.}
\end{figure*}

The $|\textrm{ENG} \rangle$ states shown in Fig. \ref{fig:Stability_NOINT} also display high stability. Oscillations are less visible because the $|\textrm{ENG} \rangle$ states are not pure $\cos(m \phi)$ wave functions: they have higher $\cos(k\phi)$ modes mixed in, representing superpositions of higher angular momentum states. The high stability of the non-interacting $|\textrm{ENG} \rangle$ states serves as a benchmark for the behavior of the interacting case, which is shown in Fig. \ref{fig:Stability_INT}. 

In the presence of self-interactions, the nonlinear term in Eq. \ref{Eq_GrossPitaevsi} couples different $\cos(k\phi)$ modes, leading to a population transfer between $\cos(k\phi)$ modes over time. This evolving mix of $\cos(k\phi)$ modes is what governs $R(t)$, resulting in lower values compared to the non-interacting case. However, good stability is maintained, with $R(t)$ always well above $90\%$. In Appendix B, we study in detail the time evolution of the decomposition of the wave function on the basis of $\cos(k\phi)$ modes. We note that the coupling between angular momentum states has also been studied in \cite{Pelegri_UnbalancedVortices} in the case of imbalanced superpositions. 

The same coupling of modes affects the evolution of the interacting $|\textrm{OAM}\rangle$ state, whose autocorrelation is reported for comparison in Appendix C. The main features of this dynamics are well described by a two-state analytical model that we develop in Appendix D.

The insets of Fig. \ref{fig:Stability_INT} are images of the atomic density at three frames separated by 0.1~s for $N=10^4$ and $m=3,9$. The population of higher modes is evident in the $m=3$ case. The images also show that the position and number of nodes remain constant. The stability of the node positions was further confirmed by a more detailed analysis of the images (not shown here).

We recall that, as proposed by \cite{Dowling_MatterWaveGyroscopy}, $|\textrm{OAM} \rangle$ states can be used to sense rotations due to the Sagnac effect. In a reference frame rotating with angular frequency $\Omega$, the states $|m \rangle$ and $|-m \rangle$ accumulate opposite $m\Omega t$ phases over time. Hence the $|\textrm{OAM} \rangle$ state evolves as: 
\begin{equation*}
    |m \rangle + |-m \rangle \Rightarrow e^{\textrm{i}m\Omega t} |m \rangle + e^{-\textrm{i}m\Omega t} |-m \rangle.
\end{equation*} 
The resulting phase difference $2m\Omega t$ leads to a precession $\Omega$ of the nodes, which can be measured by taking images of the cloud \cite{Dowling_MatterWaveGyroscopy}. Therefore the stability of the nodes shown in this section suggests that also the $|\textrm{ENG} \rangle$ states are good candidates for rotation sensing.

\section{\label{sec:conclusions}conclusions and outlook}
In this paper, we proposed a new and simple method for creating persistent current superpositions in a Bose-Einstein condensate. It offers high transfer efficiency and its experimental realization is feasible with existing light sculpting techniques. We demonstrated numerically that, even in presence of self-interactions, our engineered state achieves very high fidelity with the target state and that it remains stable in time. Our protocol is very general and can be used to engineer the wave function in an arbitrary way. For instance, it will be possible to extend it to imbalanced superpositions of persistent currents \cite{Pelegri_UnbalancedVortices}. 

Our protocol has been presented in the mean-field regime, valid for weak interactions, which produces a product state of $N$ atoms, similarly to \cite{Dowling_OAM, Pelegri_UnbalancedVortices}. In future, it would be interesting to move from the mean-field theory to the strongly-correlated regime, where strong interactions work to produce entangled states \cite{Hallwood_2006,PhysRevResearch.3.013034,briongosmerino2026}. Another interesting direction will be the extension from a single ring to coupled rings, where interesting dynamics has been studied \cite{Chaika_atomtronics,PhysRevA.96.013620}.

In the context of a linear trap, the numerical implementation in Fig.~\ref{fig:Scheme_BarrierPI} shows the transfer to a specific excited state of the trap. This can be generalized to any motional state, including
superpositions of excited states. Such a high degree of control is of fundamental importance for quantum information schemes in which information is encoded in external degrees of freedom. This will be a topic for future work.

\begin{acknowledgments}
We are grateful to Alexander Samson and to Karen Craigie for their simulations demonstrating the feasibility of the wave function engineering scheme. This project was funded by UKRI grant EP/X030369/1 "Matter-Wave Interferometers".
\end{acknowledgments}

\section{\label{sec:appendix}appendices}

\subsection{\label{wave function_distortion} The $|\textrm{ENG} \rangle$ wave function}

\begin{figure}[hbt!]
\includegraphics[scale=0.35]{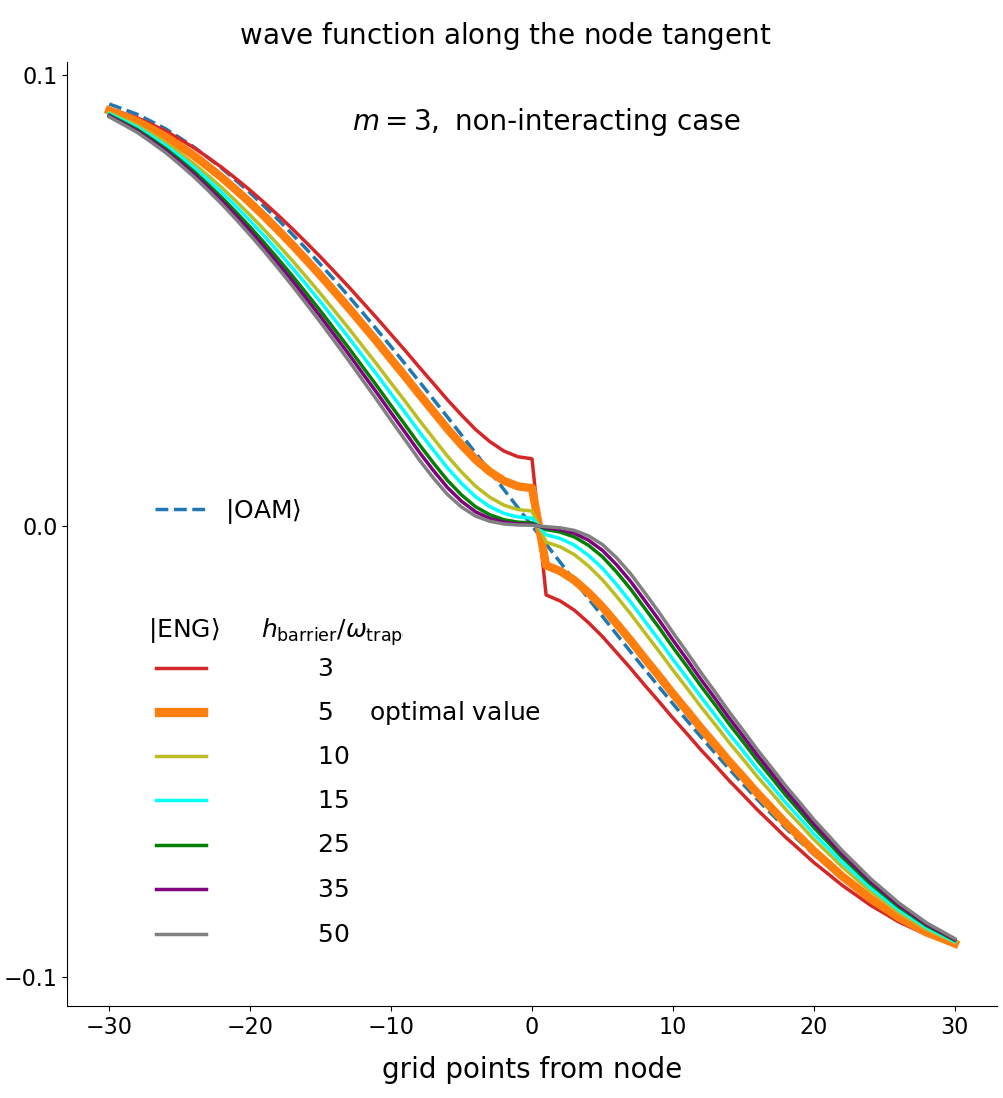}
\caption{\label{fig:Displacement_NOINT_m3} 
$|\textrm{ENG} \rangle$ wave function profiles in the vicinity of a barrier for $m=3$ and no self-interactions, covering the same range of $h_{\textrm{barrier}}/\omega_{\textrm{trap}}$ as in Fig. \ref{fig:Fidelity0}. We plot the wave function along the tangent to the ring, which allows easier computation while providing sufficient accuracy.}
\end{figure}

The behavior of the $|\textrm{ENG} \rangle$ wave function in the vicinity of a barrier helps to understand the fidelity results. Fig. \ref{fig:Displacement_NOINT_m3} shows the $m=3$ non-interacting $|\textrm{ENG} \rangle$ state immediately after phase imprint, when the wave function is real. The wave function of the corresponding $|\textrm{OAM} \rangle$ is shown for comparison. We see that for a low height $h_{\textrm{barrier}}/\omega_{\textrm{trap}}=3$, the discrepancy between the two wave functions is due to the finite value of $|\textrm{ENG} \rangle$ at the barrier. At high barriers such as $h_{\textrm{barrier}}/\omega_{\textrm{trap}}\ge 10$, the discrepancy is due to the barrier displacing the $|\textrm{ENG} \rangle$ wave function. Visual inspection confirms that $h_{\textrm{barrier}}/\omega_{\textrm{trap}}=5$ minimizes the discrepancy, in agreement with the optimal $h_{\textrm{barrier}}/\omega_{\textrm{trap}}$ value found in Fig. \ref{fig:Fidelity0}. 

With self-interactions and for increasing $m$, the distortions become more accentuated and their effects reach further away from the node (affecting the optimal value of $h_{\textrm{barrier}}$), but the general results are similar. 

\subsection{\label{decomposition} Decomposition on the cosine basis}

We gain insight into the time evolution in presence of interactions by decomposing the wave function onto an orthogonal basis. As basis, one could choose the single currents $|k \rangle=\frac{1}{\sqrt{2\pi}}f(r)e^{\textrm{i}k\phi}$ or the current superposition modes $|k_{\pm} \rangle =\frac{1}{\sqrt{\pi}}f(r)\cos(k\phi)$, which we refer to as the cosine basis. Both choices provide useful information, but we prefer the latter because the conservation of angular momentum dictates that only these modes are populated during the time evolution (given that we start from a state of no net angular momentum). 

We compute the overlaps of the modes $|k_{\pm} \rangle$ with the evolving state $|\psi(t) \rangle$ for $m=3$ and $N=10^3$. For the $|\textrm{ENG} \rangle$ case, the overlaps squared $|\langle k_{\pm} | \psi(t) \rangle|^2$ are shown in Fig. \ref{fig:Decomposition_ENG}. The corresponding $|\textrm{OAM} \rangle$ case is discussed later in Appendix~C (Fig. \ref{fig:Decomposition_OAM}). 

The numerical results show that only the modes with odd $k$ are populated over time. This can be understood with a parity argument. Our Hamiltonian is invariant with respect to parity, i.e. with respect to the transformation $x\rightarrow -x$, and with respect to the transformation $y\rightarrow -y$. This means that if a state at $t=0$ has a given parity, the evolved state must maintain that parity. Hence only the modes $|k_{\pm} \rangle$ with the same parity as the initial state are populated. For a current superposition with $m=3$, these are the modes with odd~$k$.

Fig. \ref{fig:Decomposition_ENG} shows that modes with higher $k$ already have a small population at $t=0$ for the $|\textrm{ENG} \rangle$ state. This is because the $|\textrm{ENG} \rangle$ state is not a pure $|3_{\pm} \rangle$ wave function and has higher modes mixed in to begin with. Then, at $t>0$, we observe that there is only one predominant higher mode, that is the $|9_{\pm} \rangle$. This and the corresponding results for the $|\textrm{OAM}\rangle$ state of Appendix C suggest to describe the system with a two-state model, which we report in Appendix D.

\begin{figure}[h!]
\includegraphics[scale=0.23]{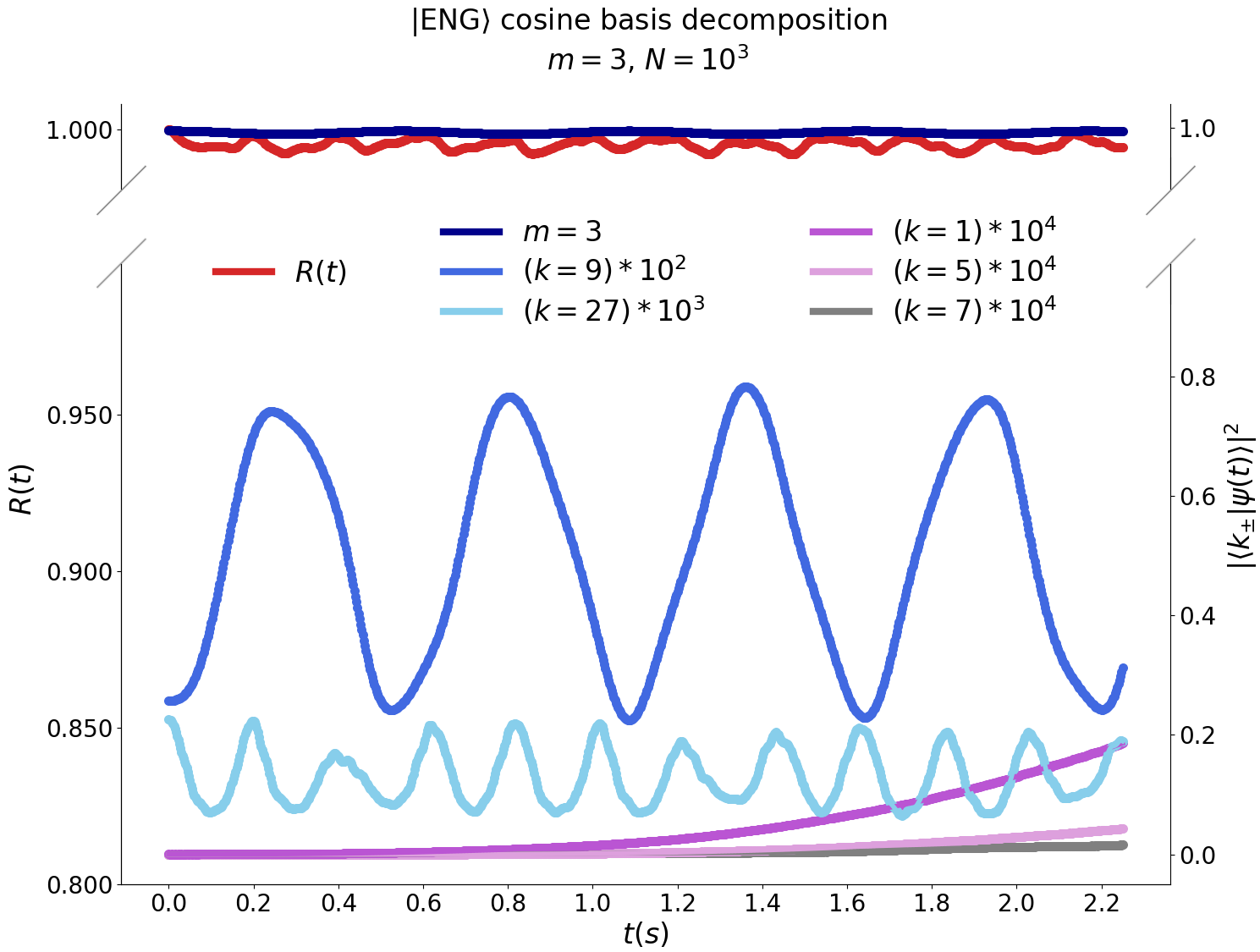}
\caption{\label{fig:Decomposition_ENG} Decomposition of the interacting $| \textrm{ENG} \rangle$ state on the cosine basis $|k_{\pm}\rangle$ for $k=1,3,5,7,9,27$. $R(t)$ is reported from Fig. \ref{fig:Stability_INT} for reference.}
\end{figure}

\subsection{\label{OAM_state} The interacting $| \textrm{OAM} \rangle$ state}

\begin{figure}[h!]
\includegraphics[scale=0.25]{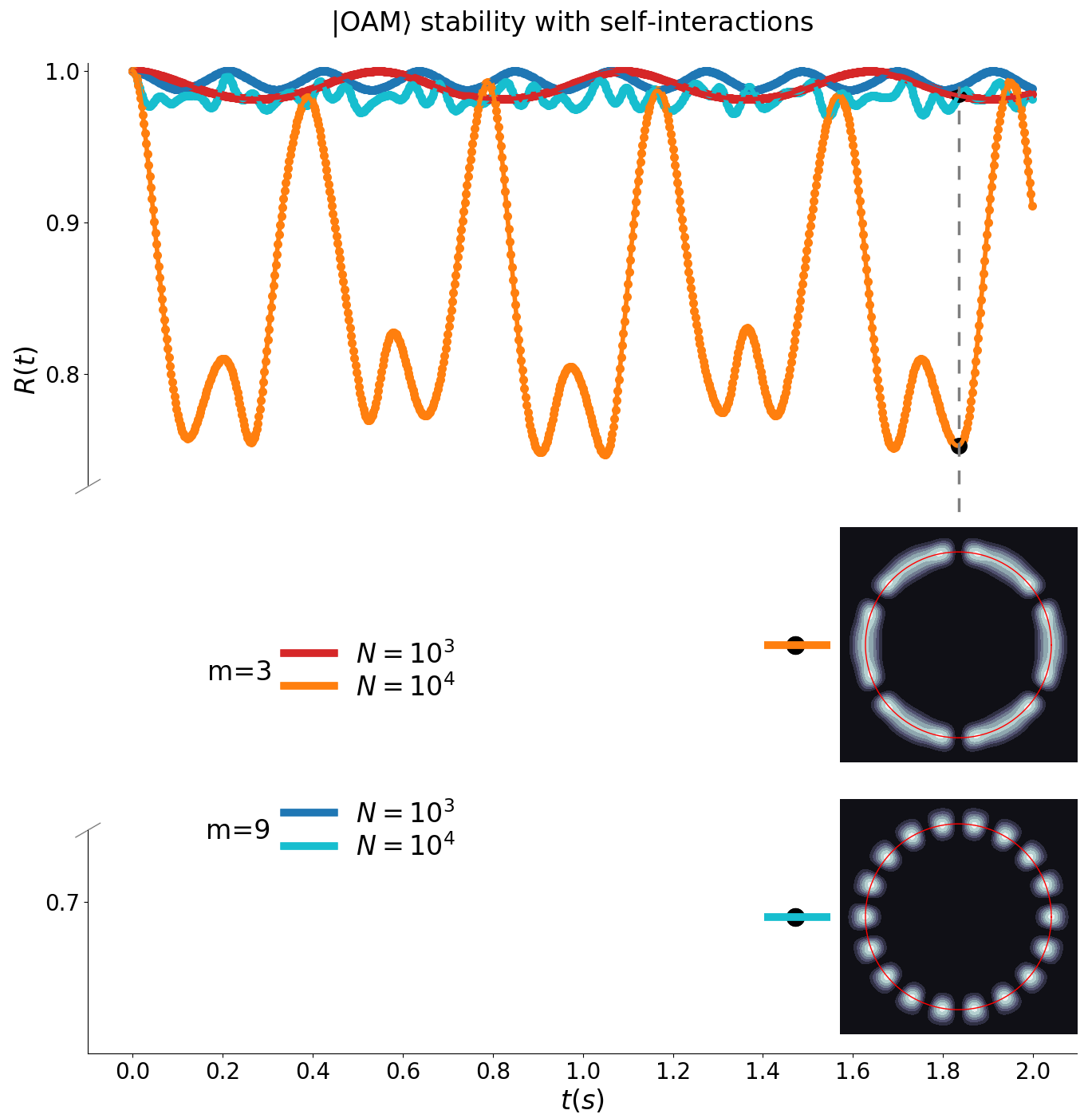}
\caption{\label{fig:Stability_OAM} Square modulus of the autocorrelation function for the interacting $| \textrm{OAM} \rangle$.}
\end{figure}

For the purpose of comparison with the $| \textrm{ENG} \rangle$ state, Fig. \ref{fig:Stability_OAM} shows the time evolution of the  $| \textrm{OAM} \rangle$ state for the same parameters as in Fig. \ref{fig:Stability_INT}. With self-interactions, the $| \textrm{OAM} \rangle$ states are no longer stationary because the nonlinear term of Eq. \ref{Eq_GrossPitaevsi} couples different $|k_{\pm} \rangle$ modes. As for the $| \textrm{ENG} \rangle$ state, this leads to oscillations of the populations of the modes over time.

The most notable result in Fig. \ref{fig:Stability_OAM} is for the case $m=3$, $N=10^4$, where $R(t)$ reaches levels as low as $0.75$. The inset shows the atomic density at a time of minimum $R(t)$, which is clearly different from the $\cos^2(m \phi)$ lobes of the $| \textrm{OAM} \rangle$ state at $t=0$. The lobes undergo an evolution qualitatively similar to that observed in Fig.~\ref{fig:Stability_INT} for the corresponding $| \textrm{ENG} \rangle$ state. However the $| \textrm{ENG} \rangle$ state is considerably more stable, with $R(t)$ remaining well above 0.90. The lobes of the $| \textrm{ENG} \rangle$ state at $t=0$ are by construction wider than $\cos^2(m \phi)$, because they are created from the ground state of the "ring + barriers" potential in presence of repulsive interactions. This is what leads to better stability of the $| \textrm{ENG} \rangle$ state over time.

In Fig. \ref{fig:Stability_OAM} we also see that the oscillations for $m=9$, $N=10^4$ are much less pronounced than for $m=3$, $N=10^4$. The reason is as follows, based on the analytical two-state model in Appendix D. In the $m=3$ case, the predominant higher mode populated during the evolution is the $|9_{\pm} \rangle$, whereas in the $m=9$ case it is the $|27_{\pm} \rangle$. In general, the difference in chemical potential between two modes is equal to the difference in their kinetic energy, which itself is proportional to the square of the angular momentum (see Eq. \ref{eq:delta_mu_m_squared}). Hence we see that for the $m=9$ case, where the predominant higher mode is the $|27_{\pm} \rangle$, there is a much larger difference in chemical potential between the two modes, compared to the $m=3$ case where the predominant higher mode is the $|9_{\pm} \rangle$ ($27^2-9^2\gg9^2-3^2$). It is then shown in the two-state model that for a given $m$, the population of the predominant higher mode is inversely proportional to this chemical potential difference (see Eq. \ref{eq:amplitudec2} for the amplitude of the oscillation of the higher mode). This results in a much smaller population of the higher mode in the $m=9$ case, hence a smaller excursion in $R(t)$.

\begin{figure}[h!]
\includegraphics[scale=0.23]{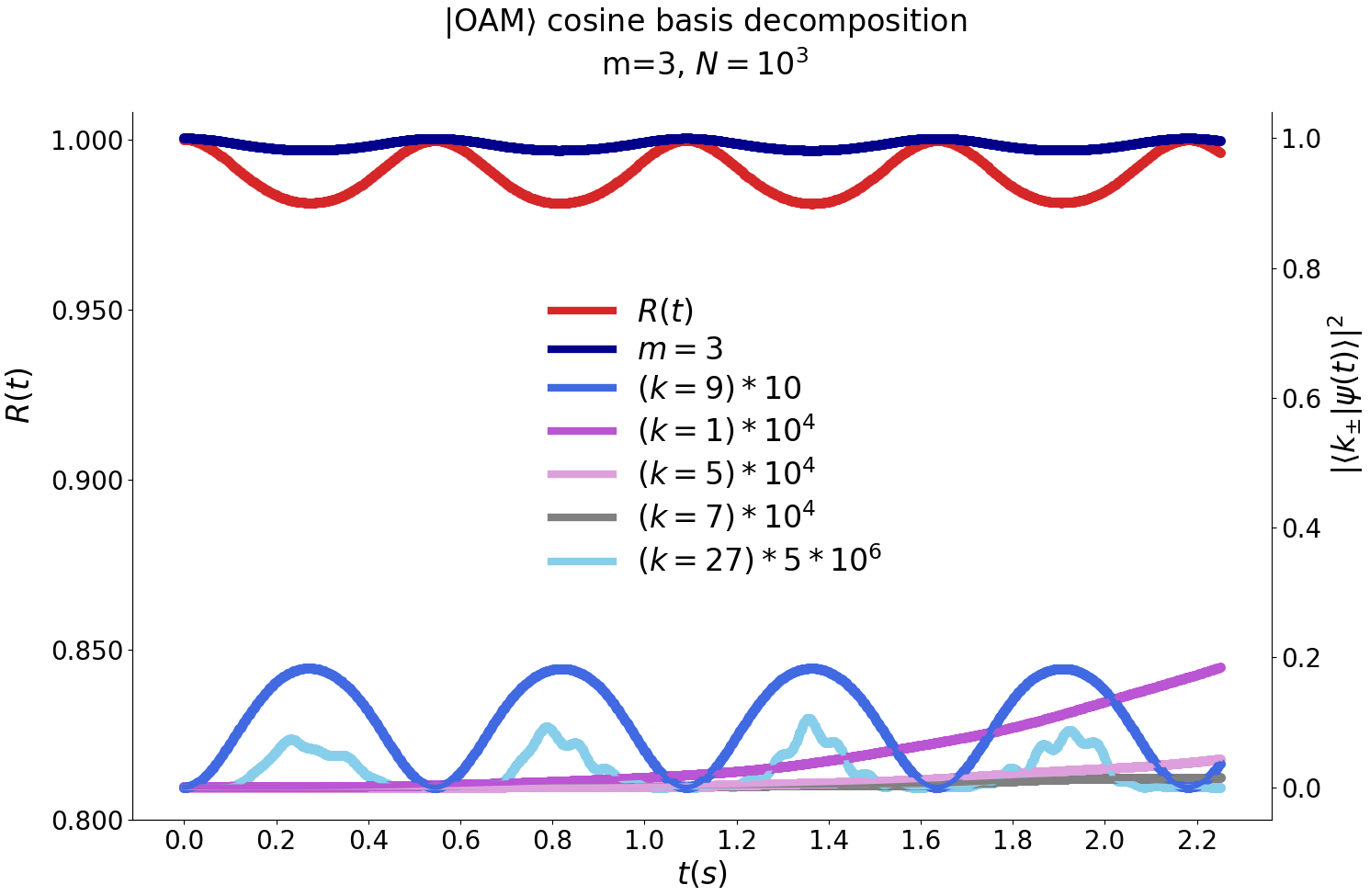}
\caption{\label{fig:Decomposition_OAM} Decomposition of the interacting $| \textrm{OAM} \rangle$ state on the cosine basis $|k_{\pm}\rangle$ for $k=1,3,5,7,9,27$. $R(t)$ is reported from Fig. \ref{fig:Stability_OAM} for reference.}
\end{figure}

Fig. \ref{fig:Decomposition_OAM} shows the cosine basis decomposition of the $| \textrm{OAM} \rangle$ state with $m=3$, $N=10^3$. Differently from $| \textrm{ENG} \rangle$, at $t=0$ the $| \textrm{OAM} \rangle$ state coincides with $|3_{\pm} \rangle$ by construction, hence the other modes are only populated in the subsequent evolution. The fact that $| \textrm{OAM} \rangle$ does not have higher modes initially mixed in leads to a cleaner characterization of the dynamics of the higher modes.

\subsection{\label{analytical} Analytical two-state model}

We develop an analytical model for the evolution of a two-state system $\psi(t) = c_1(t)\psi_1 +c_2(t)\psi_2$ in the case that one of the two modes is much less populated compared to the other: $|c_2|^2 \ll |c_1|^2$. This condition is equivalent to assuming a small self-interaction because the excitation of other modes in the condensate is due to atomic interactions, expressed through the mean-field term of Eq.~\ref{Eq_GrossPitaevsi}. 

Under these assumptions, Eq. \ref{Eq_GrossPitaevsi} can be effectively linearized and an approximated equation for $c_2(t)$ can be derived. This simple model captures the main results that describe the stability of the $|\textrm{ENG} \rangle$ and $|\textrm{OAM} \rangle$ states, depicted by the decomposition results of the previous sections. There, $\psi_1$ is represented by $|3_{\pm} \rangle$, the target mode, while the higher mode $\psi_2$ is represented by $|9_{\pm} \rangle$. In fact, our model \textit{predicts} that the higher mode must have angular momentum equal to three times that of the main mode. Moreover, the model obtains an evolution of the higher mode population $|c_2(t)|^2$ which agrees with the numerical results for $|9_{\pm} \rangle$ in Figs.~\ref{fig:Decomposition_ENG}~and~\ref{fig:Decomposition_OAM}.

For convenience of notation, in this section we drop the Dirac notation for the wave functions.
Consistently with the cosine basis used earlier, the normalized wave functions of the two modes are:
\begin{align*}
    & \psi_j = \frac{1}{\sqrt{\pi}}f(r)\cos(m_j\phi) 
\end{align*}
with $j=1,2$ and $f(r)$ such that $\int f^2(r)rdr = 1$. 

In the following, we explicitly expand the nonlinear term $g_{2D}|\psi|^2\psi$ of Eq. \ref{Eq_GrossPitaevsi}, while for the kinetic and potential energy of single modes we get (in dimensionless units):
\begin{align*}
   & (-\frac{1}{2}\nabla^2+V+g_{2D}|\psi|^2)\psi_j \approx (-\frac{1}{2}\nabla^2+V)\psi_j \approx \mu_{j}\psi_{j}
\end{align*}
with $j=1,2$, where $\mu_{j}$ are the chemical potentials for the $\psi_j$ modes. This is valid in the limit of weak interactions, because the contribution of the interaction energy to the chemical potential can be neglected and the $\psi_j$ modes are approximately equal to their non-interacting counterparts. (We note that the effect of repulsive interaction on the $\psi_j$ modes is only a small increase in the width of the radial part $f(r)$.)

Neglecting higher-order terms in $c_2$, the expansion of the nonlinear term of Eq. \ref{Eq_GrossPitaevsi} yields:
\begin{gather*}
    |\psi|^2\psi \approx \\
    \approx |c_1|^2c_1\psi_1^3 + |c_1|^2c_2\psi_1^2\psi_2 + 
    |c_1|^2c_2\psi_1^2\psi_2 + 
    c_1^2c_2^*\psi_1^2\psi_2
\end{gather*}
Performing the braket with $\psi_1$ and $\psi_2$ and integrating in space, we obtain the differential equations for $c_1$ and $c_2$ respectively. We obtain the following contributions of the nonlinear term for the equation in $\dot{c_1}$:
\begin{gather}\label{non_linearterm_eq1}
    \int_S \psi_1|\psi|^2\psi rdrd\phi = \\
    = \frac{U}{\pi^2} \bigg[ |c_1|^2c_1\int_0^{2\pi} \cos^4(m_1\phi) d\phi + \nonumber\\
    + (2|c_1|^2c_2 + c_1^2c_2^*)\int_0^{2\pi}\cos^3(m_1\phi)\cos(m_2\phi) d\phi \bigg] \nonumber
\end{gather}
and for the equation in $\dot{c_2}$:
\begin{gather}\label{non_linearterm_eq2}
    \int_S \psi_2|\psi|^2\psi rdrd\phi = \\
    = \frac{U}{\pi^2} \bigg[ |c_1|^2c_1\int_0^{2\pi} \cos^3(m_1\phi)\cos(m_2\phi) d\phi + \nonumber\\
    + (2|c_1|^2c_2 + c_1^2c_2^*)\int_0^{2\pi} \cos^2(m_1\phi)\cos^2(m_2\phi) d\phi \bigg] \nonumber
\end{gather}
where $U=g_{2D}\int_0^{\infty}f^4(r)rdr$ includes the radial contribution to the integral.

This formulation makes immediately evident the selection rule consistent with a two-mode system in which one mode is much less populated than the other. Considering that $\cos^3(m_1)=(\cos(3m_1)+3\cos(m_1))/4$, we find that the coupling is different from zero only if $m_2=3m_1$. In our case $m_1=3$, so it can be coupled only with $m_2=9$. This is indeed what we found earlier in the numerical decomposition in Figs.~\ref{fig:Decomposition_ENG}~and~\ref{fig:Decomposition_OAM}, where the predominant higher mode is the $|9_{\pm} \rangle$.

Performing the integrals for $m_2=3m_1$, Eqs. \ref{non_linearterm_eq1} and \ref{non_linearterm_eq2} become, respectively:
\begin{align*}
    & \int_S \psi_1|\psi|^2\psi rdrd\phi = \frac{U}{\pi} \big[ \frac{3}{4} |c_1|^2c_1 + \frac{1}{4}(2|c_1|^2c_2 + c_1^2c_2^*) \big]
\end{align*}
and
\begin{align*}
    & \int_S \psi_2|\psi|^2\psi rdrd\phi = \frac{U}{\pi} \big[ \frac{1}{4} |c_1|^2c_1 + \frac{1}{2}(2|c_1|^2c_2 + c_1^2c_2^*) \big]
\end{align*}
The final step requires recognizing that:
\begin{align*}
    & 2|c_1|^2c_2 + c_1^2c_2^* = |c_1|^2c_2 + c_1(c_1^*c_2 + (c_1^*c_2)^*) = \\
    & |c_1|^2c_2 + 2\Re(c_1^*c_2))c_1 \approx c_2 + 2\Re(c_1^*c_2))c_1
\end{align*}
Since $2\Re(c_1^*c_2) \ll |c_1|^2 \approx 1$, the two equations for $\dot{c_1}$ and $\dot{c_2}$ are:
\begin{equation*}
  \left\{
    \begin{aligned}
      & \big[\mu_1+ U\frac{3}{4\pi}[1 +\frac{2}{3}\Re(c_1^*c_2)]\big]c_1 + \frac{U}{4\pi}c_2 = \textrm{i}\dot{c_1} \\
      & [\mu_2+\frac{U}{2\pi}]c_2 + \frac{U}{4\pi}[1+4\Re(c_1^*c_2)]c_1 = \textrm{i}\dot{c_2}
    \end{aligned}
  \right.
\end{equation*}
and finally become:
\begin{equation}\label{eq:linearsystemc1c2}
  \left\{
    \begin{aligned}
      & \mu_1c_1 + \frac{U}{4\pi}c_2 = \textrm{i}\dot{c_1} \\
      & \mu_2c_2 + \frac{U}{4\pi}c_1 = \textrm{i}\dot{c_2}
    \end{aligned}
  \right.
\end{equation}
where we also neglected the correction terms in $U$ to $\mu_1$ and $\mu_2$. The two equations combine to give:
\begin{equation*}
    \begin{aligned}
      & \ddot{c_2} + \textrm{i}(\mu_1+\mu_2)\dot{c_2} - \mu_1\mu_2c_2 = 0
    \end{aligned}
\end{equation*}
which, with the initial condition $c_2(0)=0$, gives:
\begin{equation}\label{eq:c2(t)}
    \begin{aligned}
      & c_2(t) = A[e^{-\textrm{i}\mu_1t} - e^{-\textrm{i}\mu_2t}] \\
      & |c_2(t)|^2 = 2A^2[1-\cos(\Delta\mu t)]
    \end{aligned}
\end{equation}
where $\Delta\mu=\mu_1-\mu_2$ is the difference between the chemical potentials of the two modes. Hence we find that the oscillation frequency of $|c_2(t)|^2$ depends only on $\Delta\mu$. To estimate it, we note that the potential energy of the two modes is the same, hence:
\begin{equation*}
    \begin{aligned}
      & \mu_1 - \mu_2 = \langle E_1^{\textrm{KIN}}\rangle - \langle E_2^{\textrm{KIN}}\rangle
    \end{aligned}
\end{equation*}
In cylindrical coordinates:
\begin{gather}\label{eq:delta_mu_m_squared}
  \langle E_1^{\textrm{KIN}}\rangle - \langle E_2^{\textrm{KIN}}\rangle = \\
  \langle\psi_2\frac{1}{2}\nabla^2\psi_2\rangle - \langle\psi_1\frac{1}{2}\nabla^2\psi_1\rangle = \frac{1}{2\pi} \int_0^{+\infty}\frac{f^2(r)}{r}dr \int_0^{2\pi} \nonumber \\ 
  \bigg[ \cos(m_2\phi)\frac{\partial^2}{\partial^2\phi}\cos(m_2\phi) - \cos(m_1\phi)\frac{\partial^2}{\partial^2\phi}\cos(m_1\phi) \bigg]d\phi  = \nonumber \\
  = \frac{1}{2}(m_1^2-m_2^2) \int_0^{+\infty} \frac{f(r)^2}{r}dr \nonumber
\end{gather}
We obtain $\Delta\mu\approx 0.362$ with numerical integration, using a Gaussian centered at $r=r_0$ for $f(r)$, which is valid in the regime of weak interactions. This gives an oscillation period for $|c_2(t)|^2$ equal to $(2\pi/0.362)\textrm{T} \approx0.59$~s, where $\textrm{T}=1/\omega_{\textrm{trap}}$ is the conversion factor between the dimensionless units used in this appendix and the physical units used in the rest of the paper. This is in excellent agreement with the oscillation period of the $|9_{\pm} \rangle$ population found numerically in Fig. \ref{fig:Decomposition_OAM} for the $|\textrm{OAM}\rangle$ case.

Also in the $|\textrm{ENG}\rangle$ case, the oscillation period of the $|9_{\pm} \rangle$ population is well reproduced by the model, despite the oscillations being more irregular in Fig.~\ref{fig:Decomposition_ENG}. One obtains a numerical period of $\approx0.6$~s, as in the case of the $|\textrm{OAM}\rangle$ state. 

The amplitude of oscillation can be estimated using the initial condition on the derivative $\dot{c_2}(0)$. From Eqs.~\ref{eq:linearsystemc1c2}~and~\ref{eq:c2(t)}:
\begin{equation}\label{eq:amplitudec2}
    \begin{aligned}
      & A = \frac{U}{4\pi\Delta\mu}
    \end{aligned}
\end{equation}
For this we need to compute $U$, for which we can find an analytical expression. The Gaussian ground state wave function is $f(r) = (1/\mathcal{N)}e^{-(r-r_0)^2/(2a^2_\textrm{ho})}$, with $\mathcal{N}$ the normalization factor. Expressing the harmonic oscillator length $a_{\textrm{ho}}$ as a fraction of the radius $r_0$ of the trap, $a_{\textrm{ho}}=r_0/\rho$ as we did in Sec.~\ref{sec:numerical_model}, we rewrite $f(r)$ as:
\begin{equation}\label{eq:radial_ground_state}
    \begin{aligned}
      & f(r) = (1/\mathcal{N)}e^{-\rho^2(r-r_0)^2/(2r_0^2)}
    \end{aligned}
\end{equation}
with $\rho = 10$ in our case. Next, we compute the normalization factor:
\begin{gather*}
  \mathcal{N}^2 = \int_0^{+\infty} e^{-\rho^2(r-r_0)^2/r_0^2}rdr = \\
  \frac{1}{2}\frac{r_0^2}{\rho^2}\bigg[ -\rho\sqrt{\pi}\textrm{erf}\bigg(\rho-\rho\frac{r}{r_0}\bigg) - e^{-\rho^2(r_0-r)^2/r_0^2} \bigg]_0^{+\infty} \approx \frac{r_0^2\sqrt{\pi}}{\rho}
\end{gather*}
and:
\begin{gather*}
  \int_0^{+\infty} f^4(r)rdr = \frac{\rho^2}{\pi r_0^4}\int_0^{+\infty} e^{-2\rho^2(r-r_0)^2/r_0^2}rdr = \\
  \frac{1}{4\pi r_0^2}\bigg[ -\rho\sqrt{2\pi}\textrm{erf}\bigg(\rho\sqrt{2}-\rho\sqrt{2}\frac{r}{r_0}\bigg) - e^{-2\rho^2(r_0-r)^2/r_0^2} \bigg]_0^{+\infty} \approx \\
  \approx \frac{1}{4\pi r_0^2}2\rho\sqrt{2\pi} = \frac{\rho}{r_0^2\sqrt{2\pi}}
\end{gather*}
which gives: 
\begin{equation*}
    \begin{aligned}
      & U = g_{2D}\frac{\rho}{r_0^2\sqrt{2\pi}}
    \end{aligned}
\end{equation*}
Substituting this in Eq. \ref{eq:amplitudec2}, we obtain $A \approx 0.088$. This is in good agreement with the numerical result $A \approx 0.068$ for the oscillation amplitude of the $|9_{\pm} \rangle$ population in Fig.~\ref{fig:Decomposition_OAM}.

A similar type of coupling exists between $m_2=27$ and $m_1=9$. In this case $\Delta\mu \propto (27^2-9^2)$, and the same formula used to estimate the oscillation period of the $|9_{\pm} \rangle$ population gives a period of approximately $0.066$~s for the $|27_{\pm} \rangle$ population. Again, this agrees with Fig.~\ref{fig:Decomposition_OAM}. As expected, the amplitude of the $|27_{\pm} \rangle$ oscillations is modulated by the profile of the $|9_{\pm} \rangle$ oscillations. 

In conclusion, this simple two-state linearized model of the nonlinear Gross-Pitaevski Eq. \ref{Eq_GrossPitaevsi} is powerful enough to predict the selection rule for the coupling of the two modes, and to capture the dynamics of the higher mode. Specifically, the selection rule $m_2=3m_1$ is found to follow from the quadratic structure of the nonlinear term in the equation.

Figs.~\ref{fig:Decomposition_ENG}~and~\ref{fig:Decomposition_OAM} shows that the cosine decomposition includes other odd modes that do not obey the $k_2=3k_1$ rule. Modes such $k=1,5,7$ are present with amplitudes of similar order of magnitude, although much smaller than $k=9$ and with their coefficients rapidly decaying with increasing $k$ due to conservation of energy. The same analysis used in the two-state model allows to identify the selection rules that bring into the system these additional modes. Considering a four-state system, for example, we have a contribution from the nonlinear term of the form:
\begin{equation*}
    \begin{aligned}
      & \int_0^{2\pi} \cos(m_1\phi)\cos(m_2\phi)\cos(m_3\phi)\cos(m_4\phi)d\phi
    \end{aligned}
\end{equation*}
Expanding the integral, we have terms proportional to:
\begin{equation}\label{eq:4modes_part1}
    \begin{aligned}
      & \int_0^{2\pi} \cos((m_1-m_2+m_3)\phi)\cos(m_4\phi)d\phi + \\
      & +\int_0^{2\pi}\cos((m_1-m_2-m_3)\phi)\cos(m_4\phi)d\phi
    \end{aligned}
\end{equation}
and:
\begin{equation}\label{eq:4modes_part2}
    \begin{aligned}
      & \int_0^{2\pi} \cos((m_1+m_2+m_3)\phi)\cos(m_4\phi)d\phi + \\
      & +\int_0^{2\pi}\cos((m_1+m_2-m_3)\phi)\cos(m_4\phi)d\phi
    \end{aligned}
\end{equation}
With $m_1=3$ and $m_2=9$, the first and the second integral in Eq.~\ref{eq:4modes_part1} are non-vanishing respectively only if $-6+m_3=m_4$ and $-6-m_3=m_4$. They are satisfied respectively, for example, by $m_3=7$, $m_4=1$ and $m_3=1$, $m_4=-7$, where the sign is irrelevant for the argument of a cosine.
Similarly, the two integrals in Eq.~\ref{eq:4modes_part2} give, for example, the combinations $m_3=1$, $m_4=13$ and $m_3=1$, $m_4=11$. This justifies the appearance of other odd modes in the numerical simulations, with comparable amplitudes among them. We recall that even modes cannot be populated due to parity considerations, as shown in Appendix B.


\bibliography{apssamp}

\end{document}